# BI-EVENT TIMER FOR PHYSICS LAB

**Raju Baddi**
National Center for Radio Astrophysics, TIFR, Ganeshkhind P.O Bag 3, Pune University Campus, PUNE 411007, Maharashtra, INDIA; baddi@ncra.tifr.res.in

## ABSTRACT

Ubiquitously during experiments one encounters a situation where time lapse between two events has to measured. For example during the oscillations of a pendulum or a vibrating reed, the powering of a lamp and achieving of its full intensity. The powering of a relay and the closure of its contacts etc. Situations like these call for a time measuring device between two events. Hence this article describes a general Bi-Event timer that can be used in a physics lab for ubiquitous time lapse measurements during experiments. These measurements in turn can be used to interpret other parameters like velocity, acceleration etc. The timer described here is simple to build and accurate in performance. The Bi-event occurence can be applied as a signal to the inputs of the timer either on separate lines or along a single path in series as voltage pulses.

## 1. INTRODUCTION

Experiments in a physics lab(Strelkov 1987) ubiquitously involve measuring time between two events. For instance the time period between two passes of an oscillating pendulum is one such case. By making the pendulum bob block a beam of light falling on a photodiode the passage of the pendulum bob can be converted into a voltage signal. This can then be used to trigger an electronic circuit which would then give the time period between two such voltage pulses. Bi-event phenomena is also encountered in practical life like the double click of a mouse, closure of a switch and the time lapse in the response of an appliance. Human responses for instance like the flashing of a light or sounding of a bell and his depressing of a switch can also fall under Bi-event phenomena. Other cases in a physics lab can be the dropping of an object and its passage at two specific heights($h_1$, $h_2$) either in air or any other medium, passage of a marker on rotating disc etc. In directly the timer can also be used to find velocity, acceleration or time period in an experiment. This article describes a general electronic circuit that measures the time lapse between two voltage pulses. These pulses can arrive either sequentially or can arrive along separate paths. Time lapse measurements from a few 10 microseconds to 10's of seconds is possible. The measured time is displayed on a hexadecimal counter(12bit-CD4040) in the form of LEDs grouped in sets of 4. The article also gives a layout for easy building of the circuit on a general purpose circuit board.

## 2. THE BI-EVENT TIMER

Described here is an event timer which can be used to find the time between two events, like the time that elapses between powering of a relay and the closing or breaking of its contacts, time between the pressing of two switches, double click on a mouse, oscillations in a mechanical system like a pendulum or vibrating reed etc. The timer can be triggered in two modes, parallel and serial. In the parallel mode two independent terminals are provided, *viz* **Start** and **End**. When **Start** goes high(positive voltage pulse arrival) a counter is activated which stops when **End** goes high. Once these two events occur the timer is completely deactivated. It can only be reactivated by pressing the reset button **RST**. Now the time lapse between two fresh events can be measured again. In the serial mode the counter is activated on the decay of a sharp rising pulse(in digital jargon this is called falling edge triggered) at input **A** and it is deactivated on the decay of a second rising pulse at **A**. Once this has occured the timer is deactivated indefinitely and can only be reactivated for fresh measurement by depressing the **RST** button. The schematic circuit diagram of the event timer is as shown in Figure 1. The timing measurement involves using a clock generator which generates pulses with frequency in decade multiples of μs. Counting these pulses through a logic arrangement of gates such that the two events start and stop the counter pulsed through the clock generator **CLK**. The time lapse is obtained by the product of number of counts and the value of the position of the switch in micro-seconds(μs). The logic takes care to power the display of counts only after the counting is terminated through transistor Q1. This improves the carry over triggering of successive counters. It should be noted that the event timer does not provide conditioning of the input signals. It is just a timer which accepts conditioned signals. Extra arrangement for producing suitable activating signals for the timer is necessary. For example if one is interested in the time that lapses between the powering of a relay and closing of its contacts. Then the powering of the relay should send a positive going pulse to **Start** input terminal and the relay contacts upon closing should send a positive going pulse to the **End** input terminal. The intensity of illumination of the green LEDs connected to the outputs of CD4040 can be adjusted by choosing appropriate value for the Q1 base series resistor. On the other hand the gate that drives the base of Q1 can have another gate parallel with it with all the inputs/outputs connected correspondingly. This will improve the drive current of the output of the gate without a reduction in voltage. The 1MHz master clock generator for **CLK** can be any popular clock generator using inverter gates and 1MHz crystal. On the other hand a 1MHz clock module is another possibility. If a more easy BCD(Binary Coded Decimal) display is desired then this can be easily fulfilled using two CD4518 in falling edge trigger configuration.

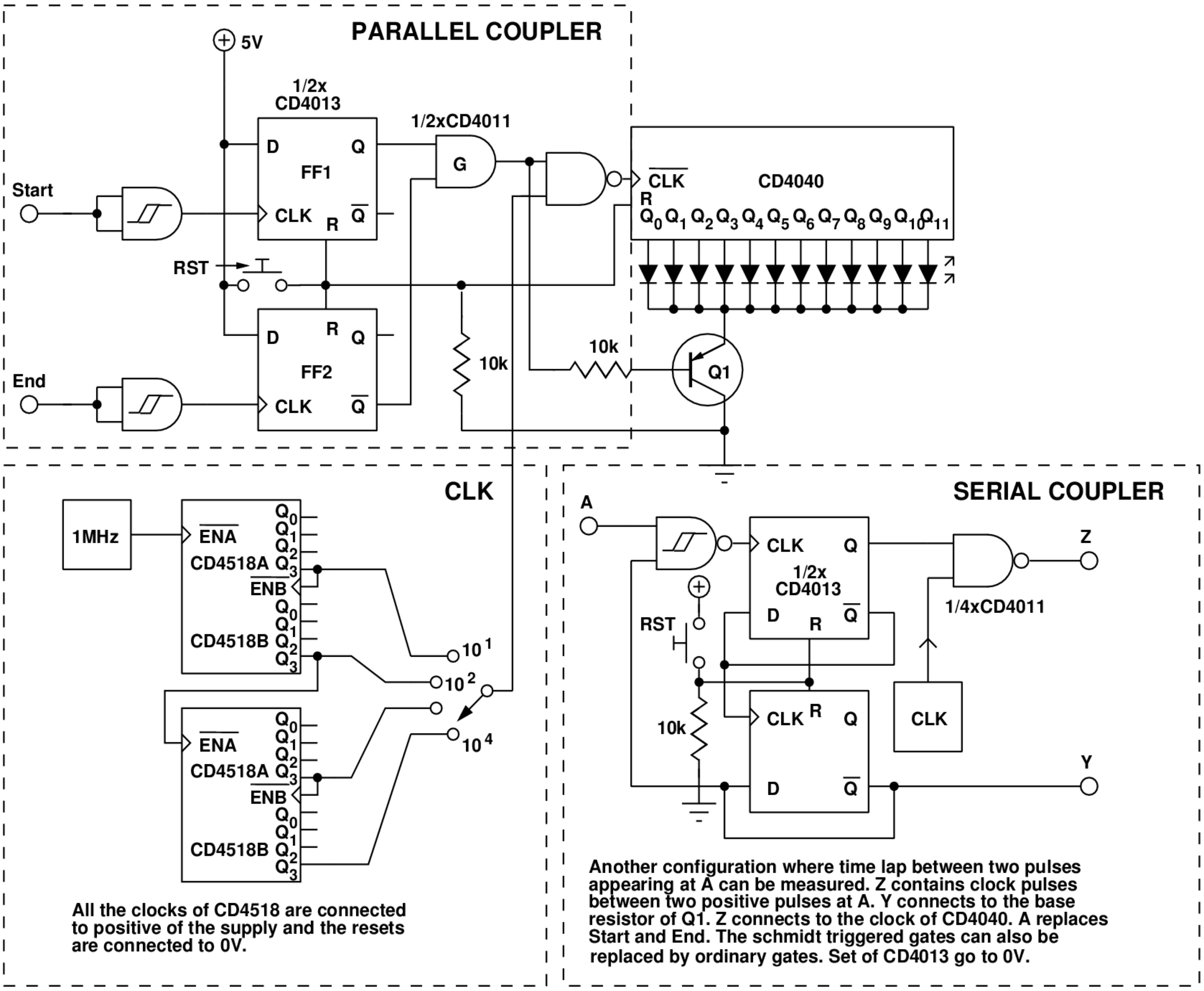

*Fig 1: Schematic circuit diagram of Bi-Event Timer. The serial coupler is shown on the right which replaces the parallel coupler already wired in. The switch is calibrated in μs. Q1 can be any suitable low power transistor(BC557). A standard NE555 monostable positive pulse instead of direct reset between the reset(RST) switch and the following circuit would further improve the performance of the Timer. For the sake of generality the schematic shows schmidt trigger gates however one can also use ordinary gates. Use of NE555 monostable(~100ms) is recommended for serial mode. Further a higher speed version can be built using TTL logic.*

**Notes:**

1) The serial coupler is a well tested configuration.
2) CD4013 CLK is assumed to be rising edge triggered.
3) CD4040 CLK is assumed to be falling edge triggered.
4) CD4518 is configured as a decade divider with falling edge trigger.
5) CD4040 is a binary counter and its counts have to be converted to decimal value before multiplication with the position value of the switch.
6) The supply voltage is 5V.
7) Ensure that the input condition is never floating.
8) The position values of the switch are in micro-second(μs).
9) NE555 is falling edge triggered. Reset switch has to be wired appropriately.

## 3. SUMMARY

This article describes a easy to build and use Bi-Event timer which can be used in numerous experiments in a physics lab to measure time lapse between two events. Since the design is simple and compact, is suitable for both the student as well as a class room. However it should be noted that the timer accepts signals which satisfy certain conditions(basically digital CMOS levels). The transducing circuit would depend on the type of experiment one is interested. For example in the case of a pendulum bob interrupting a beam of light falling on a photodiode the potential drop across the photodiode in series with a resistor can be used to activate a transistor switch which would produce the required voltage pulse at its collector(Figure 5). The proposed arrangement(Figure 2) is simple and uses LEDs grouped in sets(3) of 4 to indicate the time count. The smallest time count being 10μs and the largest time count being $10^4$μs. So the largest measureable time lapse is 4096 x $10^4$ μs. Which corresponds to a range of 40ms - 41s. If one desires a more luxurious display using 7-segment LEDs. This would involve a minor modification of the selection of counters(2 x CD4518) and using suitable segment decoders(4 x CD4511) to drive the LEDs. This however would not make the circuit layout as simple as in Figure 2.

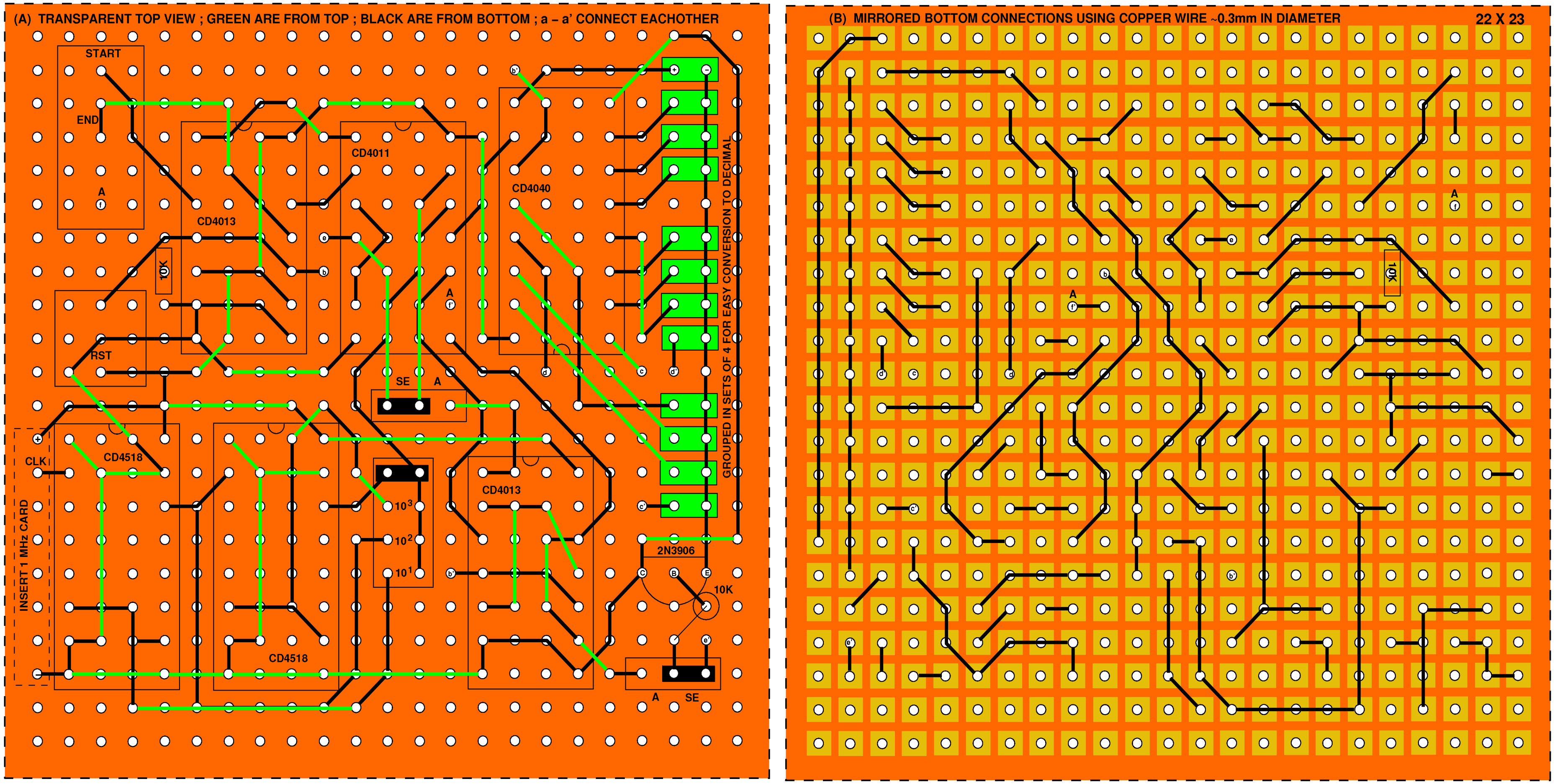

*Fig 2: Circuit layout of Bi-Event Timer. The green LEDs grouped in 4 make conversion to decimal easy. Please select suitable points to supply power to the circuit. Two sets of jumper pins allow selection between either **Start** and **End** (**SE**) or serial mode(**A**). A similar set of jumper pins allows the selection of time base frequency.*

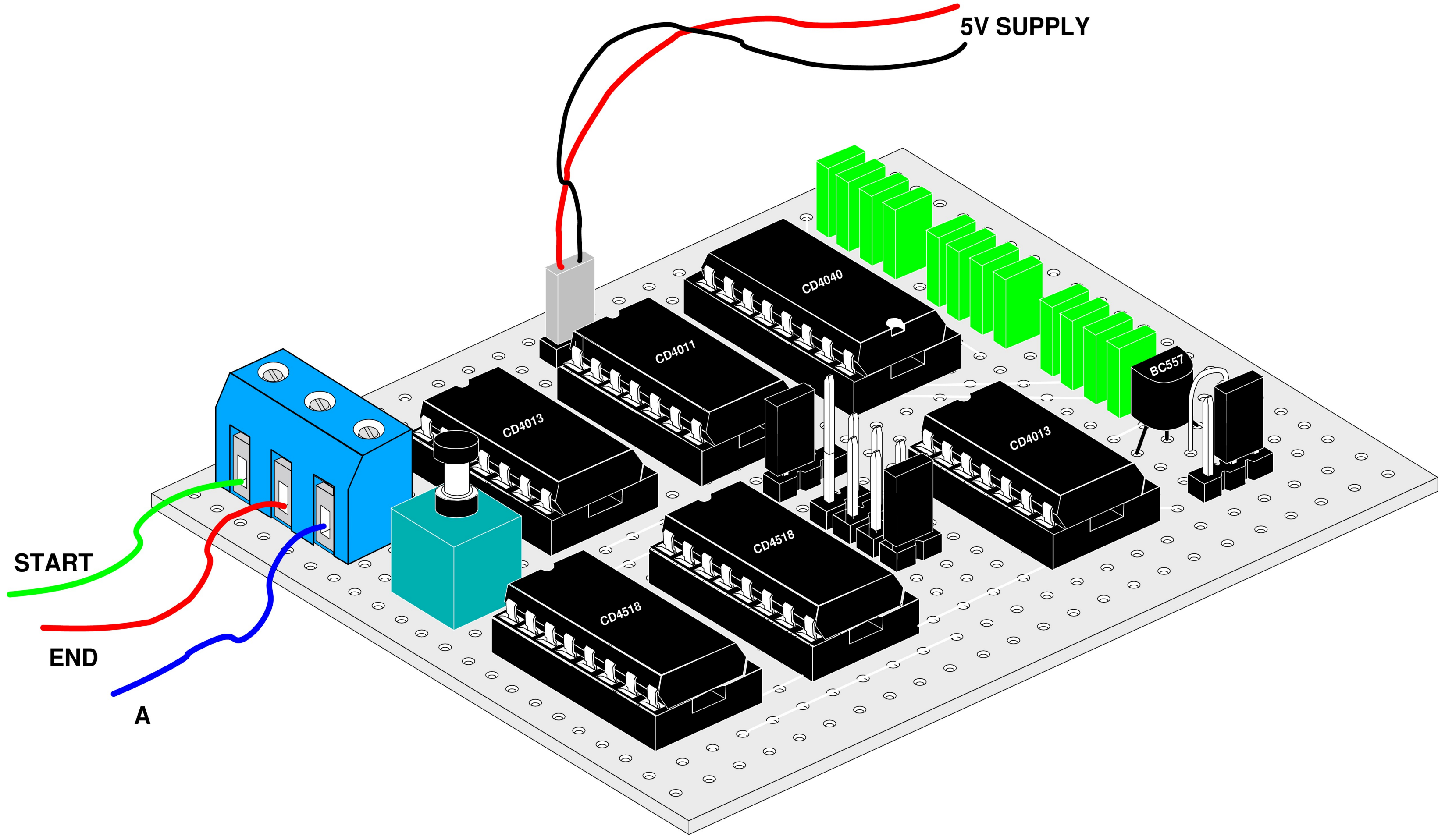

*Fig 3 : Illustrated component layout of Bi-Event Timer. Please note the illustration is just to indicate component placement and does not account for scale or accuracy. NE555 monostable has to be built on a separate circuit and connected as shown in Figure 4. The oscillator module/card can be connected after building it on a separate circuit as per the plan in Figure 1. The monostable circuit can also be accomodated as a card on the same layout(Figure 2). It should be noted that there is enough space to do this. If required the circuit size(Figure 2) can be chosen to be suitably larger, say 26 x 27, so that one can make holes at the corners to fasten the circuit to a chassis. Paint the sides of the LEDs black to make them easy readable.*

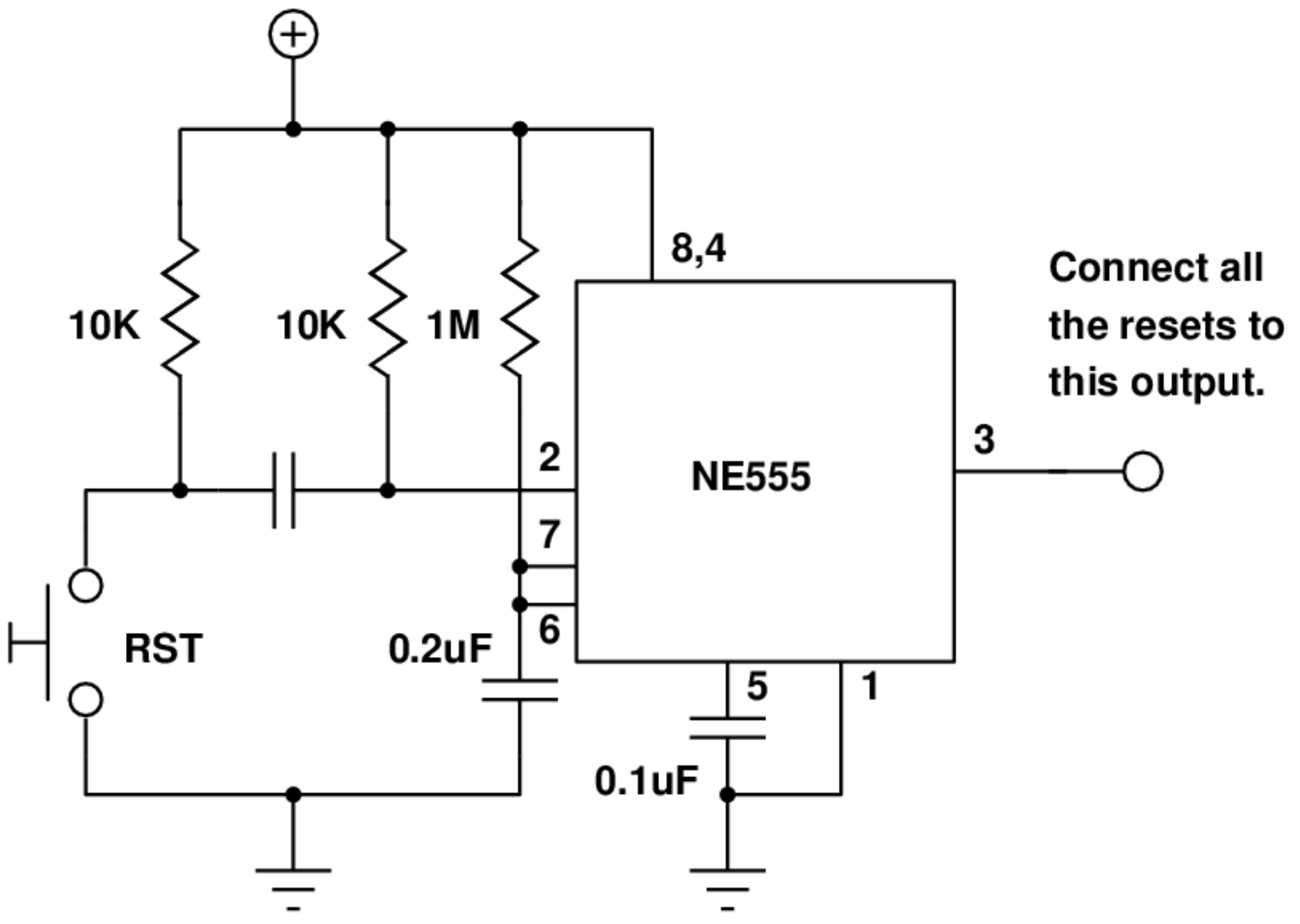

*Fig 4:  NE555 monostable which replaces the direct reset. This improves and is required during measurement of small interval serial pulses.*

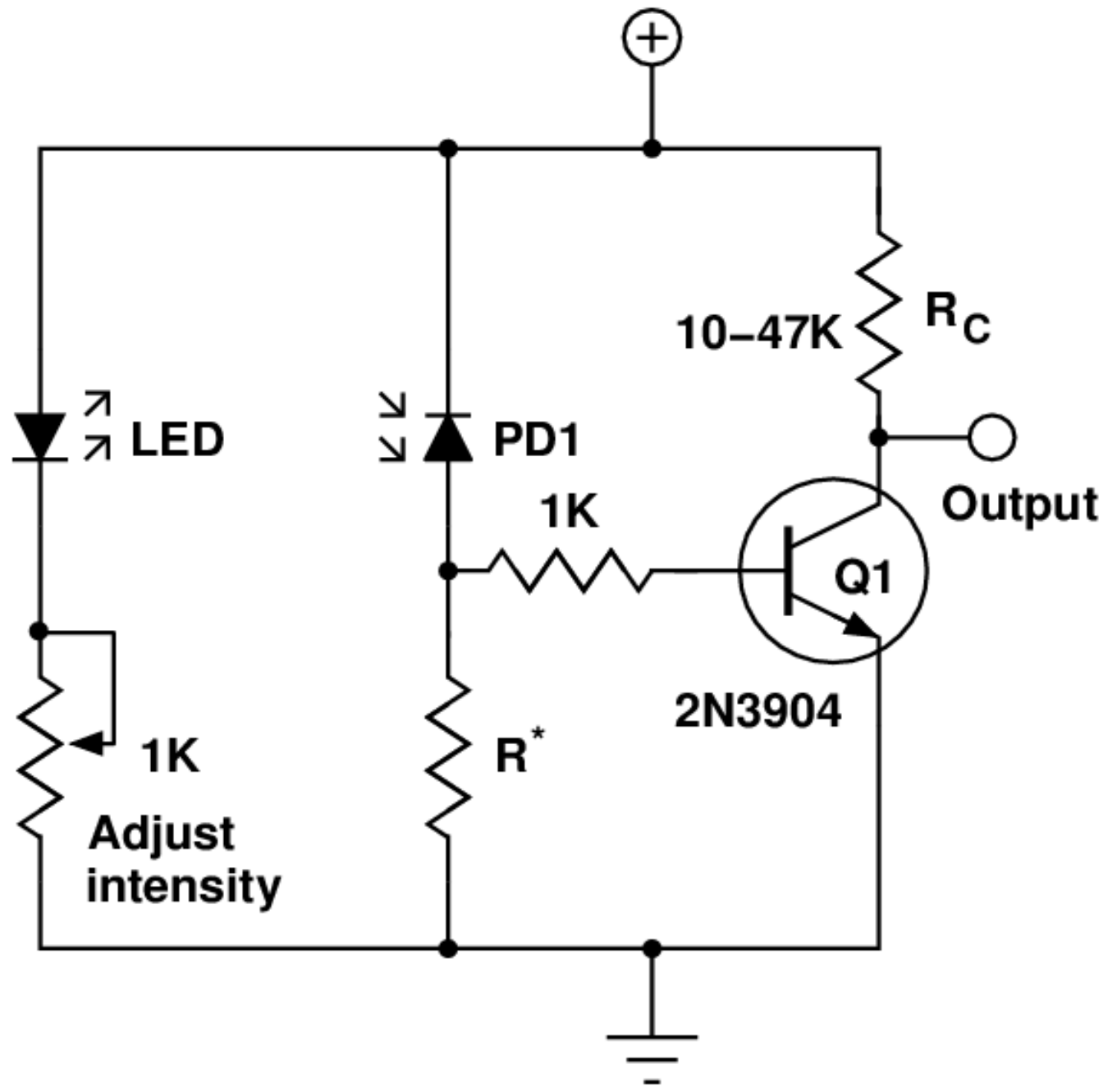

*Fig 5 :  A typical optocoupler circuit for use with Bi-Event Timer. R* can be selected by trial and error and can be on the order of 10-100KΩ depending upon the photodiode PD1. The output gives a rising positive pulse when the light falling from LED onto PD1 is interrupted.*